\documentclass{article}

\usepackage{graphicx}
\usepackage{comment}
\usepackage{caption}
\usepackage{authblk}
\usepackage[top=2cm,right=3cm,left=3cm]{geometry}

\usepackage{amssymb,amsfonts,amsmath}
\usepackage{color,soul}

\title{Scaling of phloem structure and optimality of photoassimilate transport in conifer needles} 

\author[1]{Henrik Ronellenfitsch\thanks{henrik.ronellenfitsch@ds.mpg.de}}
\author[2]{Johannes Liesche}
\author[3,4]{Kaare H. Jensen}
\author[3]{N. Michele Holbrook}
\author[2]{Alexander Schulz}
\author[1]{Eleni Katifori}

\affil[1]{Max Planck Institute for Dynamics
and Self-Organization, Am Fa\ss berg 17, 37077 G\"ottingen, Germany}
\affil[2]{Department of Plant and Environmental Sciences, 
University of Copenhagen, DK-1871 Frederiksberg C, Denmark}
\affil[3]{Department of Organismic and Evolutionary Biology, 
Harvard University, Cambridge, MA 02138, USA}
\affil[4]{Department of Physics, Technical University of Denmark, DK-2800 Kgs. Lyngby, Denmark}

\begin{document}

\maketitle
\begin{abstract} The phloem vascular system facilitates transport of energy-rich sugar and signaling molecules in plants, thus permitting long range communication within the organism and growth of non-photosynthesizing organs such as roots and fruits. The flow is driven by osmotic pressure, generated by differences in sugar concentration between distal parts of the plant. The phloem is an intricate distribution system, and many questions about its regulation and structural diversity remain unanswered. Here, we investigate the phloem structure in the simplest possible geometry: a linear leaf, found, for example, in the needles of conifer trees. We measure the phloem structure in four tree species representing a diverse set of habitats and needle sizes, from 1 cm (\textit{Picea omorika}) to 35 cm (\textit{Pinus palustris}). We show that the phloem shares common traits across these four species and find that the size of its conductive elements obeys a power law. We present a minimal model that accounts for these common traits and takes into account the transport strategy and natural constraints. This minimal model predicts a power law phloem distribution consistent with transport energy minimization, suggesting that energetics are more important than translocation speed at the leaf level.
\end{abstract}
\section{Introduction}
Plants, diverse species of algae, and other organisms acquire chemical energy in the form of carbohydrates (sugars) from the sun through photosynthesis. These photoassimilates are exported from plant leaves via the phloem vascular system to support life in distal parts of the organism. The flow is driven by a build-up of osmotic pressure in the veins, where high concentrations of sugar direct a bulk flow of sweet sap out of the leaf. The vascular network is critical to sustenance and growth; more subtly, it is also of importance as carrier of signaling molecules that integrate disparate sources of information across the organism. The mechanisms that influence phloem structure to fulfill these multiple objectives, however, remain poorly understood. 

The phloem is a complex distribution system responsible for transporting a large number of organic molecules, defensive compounds, and developmental signals by bulk fluid flow through a network of enlongated sieve element cells
which are connected to each other by porous sieve plates, effectively
forming long tubes \cite{Knoblauch2010}. 
The phloem thus serves functions analogous to a combination of the nervous- and circulatory systems of animals.
The role of the phloem in the transport of photoassimlates has been known since the 17th century \cite{Taiz2010}, but it was not until 1930 that the role of the phloem sieve element as the channel of carbohydrate transport in plants was experimentally demonstrated \cite{Schumacher1930,Mason1937,Crafts1971}. Although the primary role of phloem transport is to distribute the products of photosynthesis, it also plays a role in long-distance transmission of signals for some developmental and environmental responses. For instance, flowering is induced by transmission of a phloem-mobile hormone from the leaves to the meristem \cite{Lough2006}. Also, pathogen protection and related gene expression signals have been shown to occur through the phloem \cite{Lough2006}. 

\begin{figure}
\centering
\includegraphics[width=0.8\columnwidth]{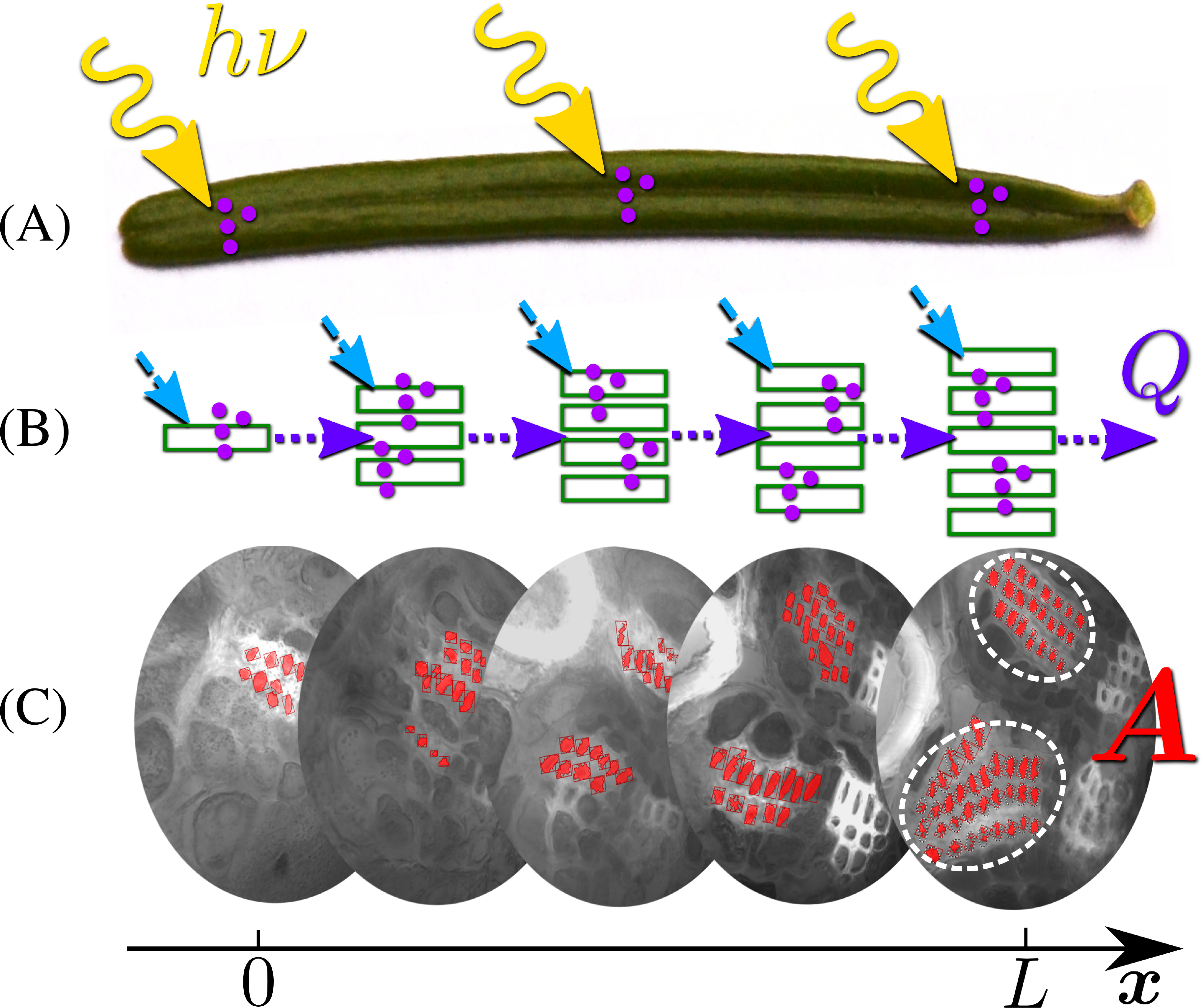}
\caption{(Colour online) Phloem geometry in a conifer needle. (A) Schematic representation of a photosynthesizing \textit{Abies nordmanniana} needle.
(B) The mechanism driving sugar export in plant leaves. Sugar molecules (purple dots) are produced via photosynthesis and diffuse into the phloem cells. The osmotic sugar solution attracts water from the surrounding tissue (blue dashed arrows) which generates a bulk flow $Q(x)$ along the needle (purple dotted arrows).
(C) Micrograph cross-sections of the phloem of an \textit{Abies nordmanniana} needle taken at distances $x=0.4, 0.8, 1.6, 2.9, 4.9\,\mathrm{mm}$
from the tip. The diameter of the circular cross-sections is 
$100\mathrm{\mu m}$.
The conductive phloem area $A(x)$ (red cells) increases with distance $x$ from the needle tip while the size of individual cells is roughly constant.
\label{fig:abies-cross-section}}
\end{figure}

Phloem flow occurs in an interconnected network of long, narrow cylindrical cells. In these cells, an energy-rich solution of sap containing $10-30\,\%$ wt sugars flows toward distal regions of the plant  \cite{Muench1930xx}. Transport is driven by the osmotic M\"unch pump \cite{Muench1930xx, Horwitz1958,Thompson2003,Pickard2009a,Jensen2011,Jensen2012}. The flow is initiated in the leaf, where sugars produced by photosynthesis accumulate in phloem cells. This induces an osmotic gradient with respect to the surrounding tissue, drawing in water into the phloem cells. On the scale of the phloem tissue, this process results in a bulk flow of sap along the major veins out of the leaf, towards regions of low osmotic potential in the plant, such as roots or fruits. M\"unch flow in a conifer needle is sketched in Fig.~\ref{fig:abies-cross-section}, where the phloem tissue is located within a single large vein near the center of the leaf cross-section plane.
Close to the tip of the needle only few sieve tubes exist to support
the flow of sap, more continually being added to the bundle as one moves 
closer to the petiole. Neighboring sieve tubes maintain 
hydraulic connections through plasmodesmata.

The driving force responsible for carbon export is the steady production of photosynthate in the leaf mesophyll located close to the leaf surface. The mechanism by which sugars accumulate in the phloem varies between species. Most plants, however,  can be roughly divided into two groups: active and passive phloem loaders \cite{Turgeon2006}. Active loaders use membrane transporters or sugar polymerization to accrue carbon in the phloem, while passive loaders rely on cell-to-cell diffusion aided by bulk flow through plasmodesmata pores \cite{Turgeon2006,Turgeon2010,Rennie2009}. Trees are predominantly passive loaders, while many herbaceous plants use active phloem loading \cite{Turgeon2006,Turgeon2010,Rennie2009,Liesche2011xx}.

%
The quantity of material exported through the phloem is generally assumed to be strongly dependent on physiological factors such as solar radiation intensity and water availability \cite{Sevanto2014, Choat2012}, but it also likely to depend on details of the leaf vascular architecture. For instance, the positioning and network structure of water-transporting xylem conduits in plant stems and leaves has been shown to play an important role in determining the efficiency of CO$_2$ uptake \cite{West1999, Savage2010, Katifori2010a,Corson2010,Zwieniecki2006a,Noblin2008a}. While the branched vein network architecture in plant leaves has been studied extensively, less is known about the functional elements. Except for a few species of grasses, which have a parallel vein geometry similar to needles \cite{Colbert1982,Russell1985,Dannenhoffer1990}, the detailed functional architecture of the phloem (i.e. the location, number, and size of conducting elements) in plant leaves remains unknown, in part due to the high sensitivity of phloem tissue to disturbances \cite{Knoblauch2014a}.

In this work, we aim to answer two basic questions. First, we ask what is the design of the phloem vascular system in conifer needles, i.e. what are its geometric and hydrodynamic properties. Second, we aim to determine whether the observed structure is consistent with energy minimization or maximizing flow rate to elucidate the selective force that influences phloem structure.
We chose conifer needles for this study in part due to their linear structure without branching veins, a geometric feature which greatly simplifies the analysis of the transport process. Moreover, conifer trees inhabit diverse environments and the vascular network has thus been subjected to a broad selective pressure.
Accordingly, we present the experimental results of phloem geometry in needles of four conifer species from a 
diverse set of habitats and needle sizes. Based on these, we develop a minimal mathematical model of sugar transport in leaves, and use a constrained optimization to derive the optimal phloem geometry in a one-dimensional leaf. Finally, we compare modeling with experimental results, and conclude by discussing the implications of our results for the study of conifers and plants in general.

\section{Results}
\subsection{Phloem structure}
\begin{figure*}
\centering
\includegraphics[width=0.8\columnwidth]{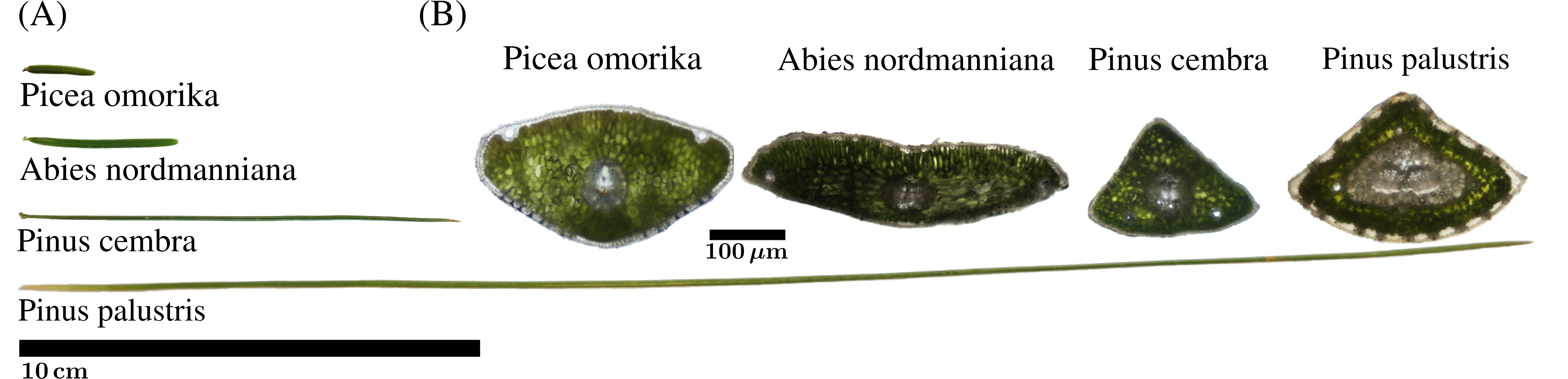}\\
\caption{(Colour online) Conifer needle characteristics. (A) Photographs of representative needle specimens of the species studied in
this work. Typical lengths range between $1\,\mathrm{cm}$ and
$30\,\mathrm{cm}$. From top to bottom, \textit{Picea omorika}, \textit{Abies nordmanniana},
\textit{Pinus cembra}, \textit{Pinus palustris}.
(B) Micrographs showing-cross sections of a typical needle specimen. The vascular tissue in
the center of each specimen (light color) is clearly discernible.
\label{fig:needles}}
\end{figure*}

We measured the phloem geometry in needles of four conifer species shown in Fig.~\ref{fig:needles}:
\textit{Abies nordmanniana}, \textit{Pinus palustris}, \textit{Pinus cembra}, and \textit{Picea omorika}. Three to six needles were sampled from each species.
The species encompass the range of typical needle sizes of conifer species, from
\textit{P. omorika} (needle length $\ell \sim 1\; \mathrm{ cm}$) to
\textit{P. palustris} ($\ell \sim 35\; \mathrm{ cm}$).
Additionally, they incorporated plants from diverse habitats
and climates ranging from \textit{P. palustris}, whose habitat are the Gulf and Atlantic
coastal plains of the United States, to the European alpine \textit{P. cembra}.  
The measurements were conducted by performing transverse sections at 10-20 positions along the length of the needle. Phloem cells were identified by the presence of a stain as described in the materials and methods section. A typical stack of images obtained this way is shown in Fig.~\ref{fig:abies-cross-section} (C).

\begin{figure}
\begin{center}
\includegraphics[width=0.8\columnwidth]{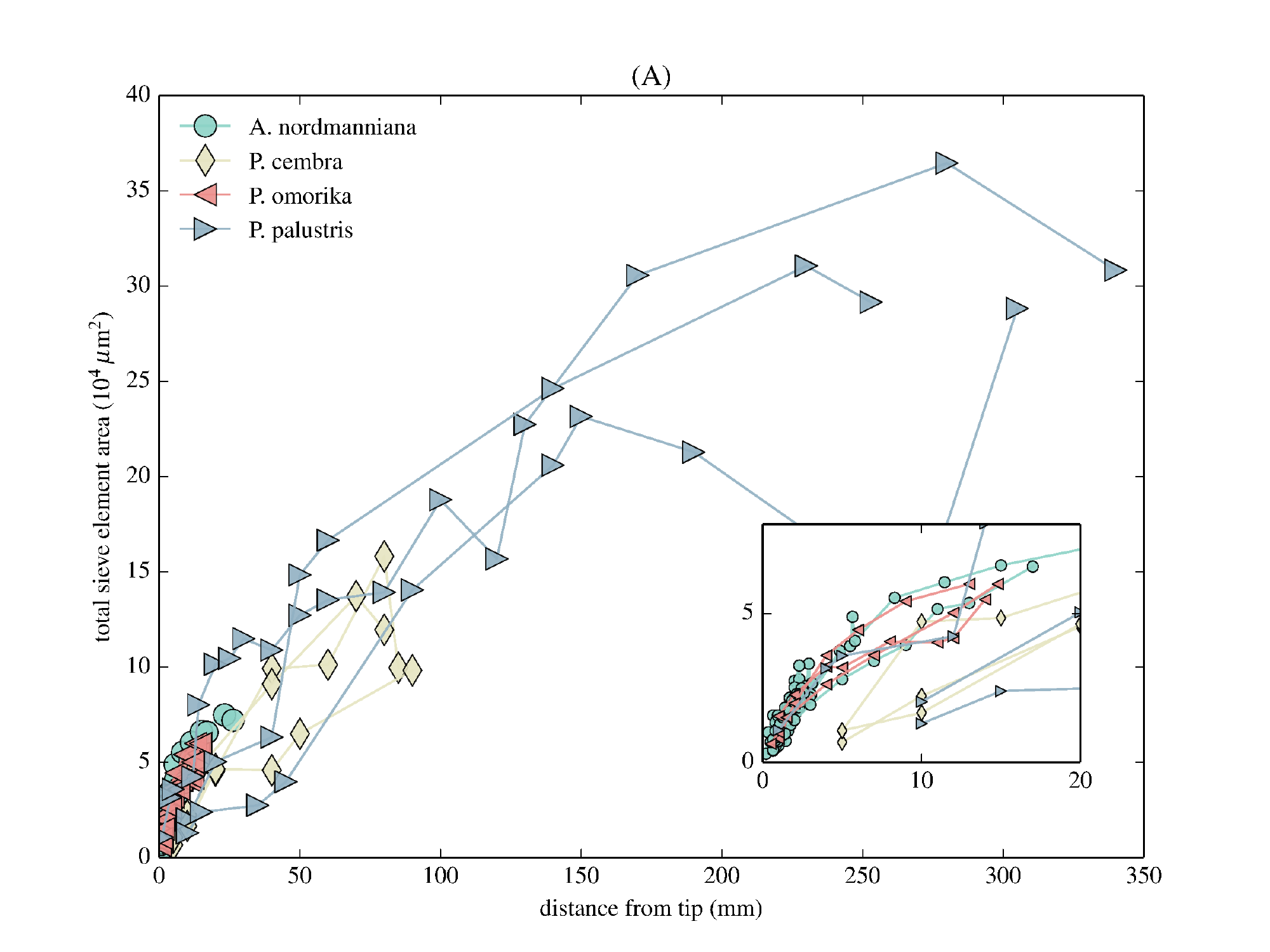}\\
\includegraphics[width=0.8\columnwidth]{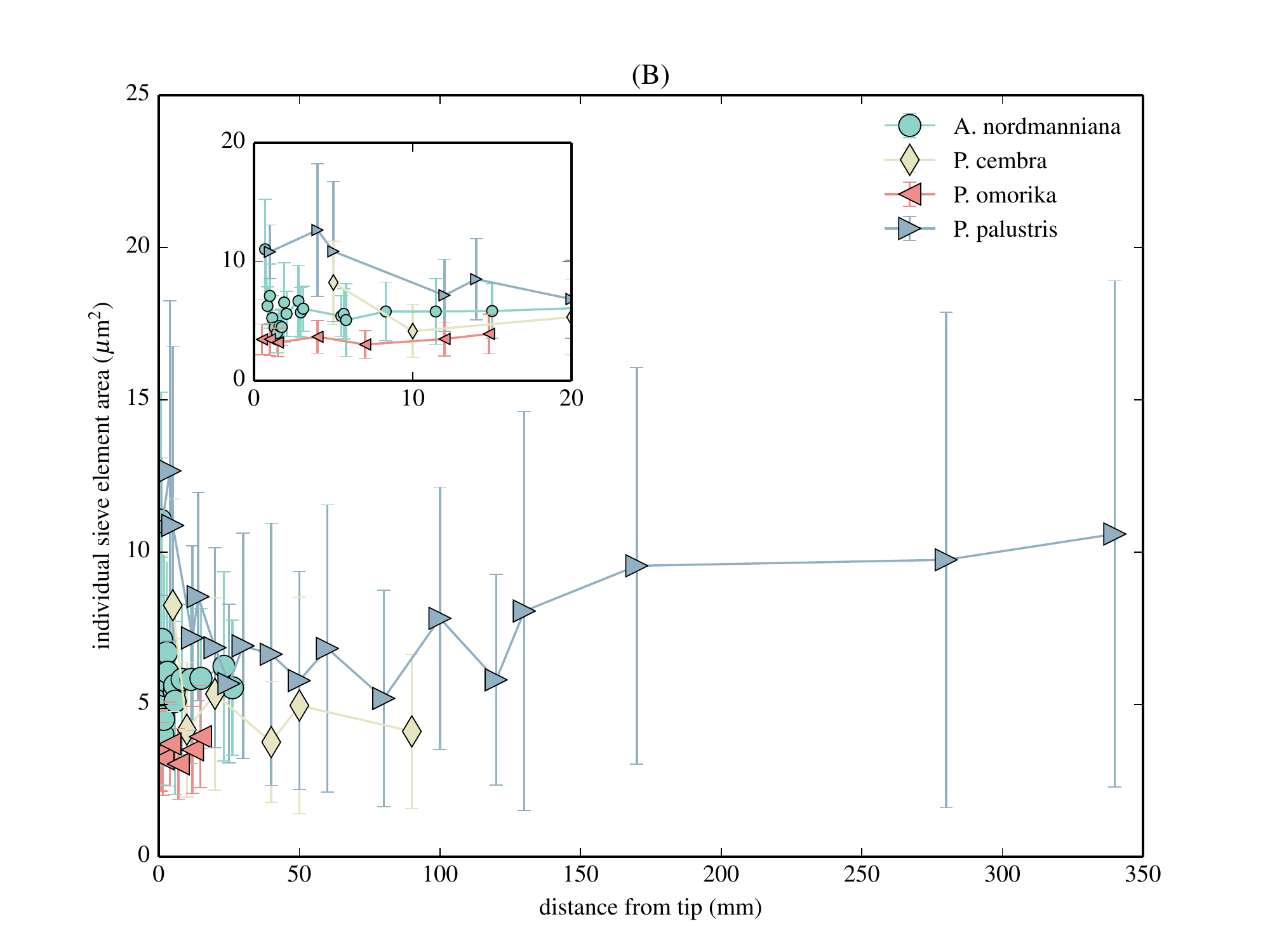}
\end{center}
\caption{(Colour online) 
Phloem area increases with distance from the needle tip. 
Total conductive phloem area $A$, shown in panel (A) 
for all analyzed needles, 
and cross-sectional area of individual sieve elements $a$, 
shown in panel (B) for one needle of each species, 
as a function of distance $x$ from the needle tip for the conifer 
species indicated in the legend (see also Fig.~\ref{fig:needles}).
Error bars in (B) correspond to one standard deviation.
The individual sieve element areas show only little variation
as a function of position, as evidenced from \emph{Pearson's}
$|r| < 0.23$ for all needles shown, 
implying that it is mainly the number of sieve elements
which contributes to hydraulic efficiency.
Data from individual needles are connected by solid lines.
Insets show details near $x=0$ for the shorter species.
\label{fig:master-a-linlin}}
\end{figure}

To quantify the phloem structure we measured the size of all sieve elements in each cross section. Starting from the tip of the needle, we typically observed an increase in the total conductive area $A$ towards the base of the leaf (Fig.~\ref{fig:master-a-linlin} (A)). 
The cross-sectional area of individual sieve elements $a$ 
(Fig.~\ref{fig:master-a-linlin} (B)), 
however, shows only minimal variation as a function of length 
(correlation to position using \emph{Pearson's} $|r| < 0.23$ for all 
needles in Fig.~\ref{fig:master-a-linlin} (B)),
implying that the main variation in transport area is driven by changes in the number of conduits $N$. When the total phloem transport area is normalized and plotted relative to the needle length $L$ on logarithmic axes (Fig.~\ref{fig:master-a-loglog}), it is seen to behave roughly as a power law $A/A(L)\sim (x/L)^\alpha$ with average exponents per species between $\alpha = 0.45$ (\textit{P. omorika}) and $\alpha=0.56$ (\textit{A. nordmanniana}), see Table 1. The number of sieve elements $N(x)\sim x^{\alpha}$ follows a similar scaling with $\alpha \approx 1/2$. Since the cross-sectional area of individual sieve elements is nearly constant, this result is to be expected.

 \subsection{Mathematical model for sugar transport in plants}
 To rationalize the observed vein structure, we develop a simple 
 model of one-dimensional sugar transport in a bundle of parallel phloem tubes based on the work of Horwitz \cite{Horwitz1958} and Thompson and Holbrook \cite{Thompson2003}.
 Sugar flow commences near the needle tip ($x=0$), where a few phloem conduits initiate the export of photoassimilates. Approaching the needle base ($x=L$), the number of conducting channels $N$ increases while the size of individual phloem tubes remains  constant.
Because the length scale at which $N$ varies is small compared to
the total length $L$ of the needle, we can approximate $N$ well by
a smooth function. We note that the precise way in which the number
of sieve tubes changes (be it by simple addition of new tubes or
branching of existing ones) has no impact on the continuum description.
Phloem loading in conifers is thought to be passive, driven by cell-to-cell diffusion across microscopic (plasmodesmata) channels \cite{Liesche2011}.  The sugar loading rate per unit length of needle $\Gamma$ is proportional to the rate of photosynthesis and to the circumference of the needle, both of which are approximately constant along the needle (see Fig.~\ref{fig:needles}). For a collection of parallel phloem tubes, conservation of sugar mass can be expressed as

\begin{align}
\frac{\mathrm d J}{\mathrm d x} = \Gamma,
\label{eqn:loading}
\end{align}
where $J(x) = Q(x)c(x)$ is sugar current with $Q(x)$ the total volume 
flow rate and $c(x)$ the sap sugar concentration. 
We further assume that at each point water enters the sieve elements 
by osmosis

\begin{align}
\frac{\mathrm dQ}{\mathrm dx} = 2\frac{L_pA}{r_0} (RT \Delta c - \Delta p),
\label{eqn:net-currents}
\end{align}
where $A(x)$ is the total conductive phloem area at the 
position $x$ (see Fig.~\ref{fig:abies-cross-section} for visualization). 
Note that $A(x) = N(x) A_0$, where $N(x)$ is the number of sieve tubes
at position $x$ and $A_0$ is the cross sectional area of a single 
sieve tube.
In Eq.~\eqref{eqn:net-currents}, $L_p$ denotes the permeability of the sieve element membrane, $R$ is the universal gas
constant, $T$ is the absolute temperature, and $r_0$ is the radius of one single sieve element.
The sugar concentration $\Delta c=c(x)-c_1$ available for driving an osmotic flow is the difference between the concentration in the sieve element $c$ and the constant osmotic concentration $c_1$ of the surrounding cells. Likewise, the pressure $\Delta p = p (x)-p_1$ is the difference between the cytoplasmic pressure $p$ and the constant pressure in neighboring cells $p_1$. For clarity we use the van't Hoff value $RT\Delta c$ for the osmotic pressure in Eq.~\eqref{eqn:net-currents}, which is valid only for dilute (ideal) solutions. At the concentrations relevant to phloem sap ($c\leq 1$ M), the error in the osmotic pressure introduced by using the van't Hoff value is $\approx10\%$ \cite{Cath2006}.
Equation \eqref{eqn:loading} may be integrated to yield $J(x)= \Gamma x$, having imposed a vanishing current at the tip.
The total export of sugar from the needle is therefore $J(L) = \Gamma L$, proportional to the loading rate $\Gamma$ and needle length $L$. The factors contributing to the energetic cost of transport include the metabolic energy required to maintain the vasculature and the power dissipated by the flow due to viscous friction. We proceed to consider how the phloem structure influences the magnitude of these contributions, and note that similar energy considerations have been used in the study of other biological transport systems \cite{Katifori2012xx,Murray01031926xx,Murray01051926xx,Zwieniecki2006a}. {For instance, Zwieniecki and co-workers derived the optimal distribution of tracheids in a pine needle that minimizes the pressure drop required to drive transpiration for a given investment in xylem conduit volume} \cite{Zwieniecki2006a}.
The phloem consists of severely reduced cells which shed most of their
organelles during maturation. It is, however, alive and relies on external
supply of metabolic energy. The rate of energy consumption by the phloem
tissue itself may be seen as an energetic maintenance cost of the
transport conduit.
Here, we assume that this energetic cost of maintaining the phloem vasculature is proportional to the conductive volume $V_0=\int_0^LA(x)\,\mathrm dx$,
or equivalently the number of phloem cells (assuming cells of roughly equal
size).
The viscous power dissipation per unit length is $\mathrm d W = -(Q\, \mathrm dp + p\, \mathrm d Q)$. To determine $W$, we note that the local pressure gradient is related to the flow speed $u=Q/A$ by Darcy's law

\begin{align}
u(x) = - \frac{k}{\mu} \frac{\mathrm d p}{\mathrm d x},
\end{align}
where $p(x)$ is the pressure, $\mu$ is the viscosity of phloem sap 
(typically around
5 times that of water \cite{Jensen2013xx}) and 
$k=A_0/8\pi$ is a geometric constant which solely depends on the
cross sectional area $A_0$ of single sieve elements.
Integrating Eq.~\eqref{eqn:loading} leads to $J=uAc=\Gamma x$, and thus the local pressure gradient is given by

\begin{align}
\frac{\mathrm dp}{\mathrm d x} = -
  \frac{\mu}{k} \frac{\Gamma}{c(x)} \frac{x}{A(x)}.
\end{align}
Analysis of the coupled system in Eqns.~\eqref{eqn:loading} and \eqref{eqn:net-currents} have shown that the concentration $c(x)$ is approximately constant along the needle $c(x) \approx c_0$ \cite{Thompson2003, Jensen2012} (see also the Supplementary Information). Integrating the differential relation for $W$ using the above set of approximations,
the total power dissipation is

\begin{align}
W = Q(L) \Delta p = \frac{\mu }{k} \frac{\Gamma^2}{c_0^2}
  \int_0^L \mathrm dx \frac{x}{A(x)}.
\label{eqn:functional}
\end{align}

\begin{figure}
\centering
\includegraphics[width=0.8\columnwidth]{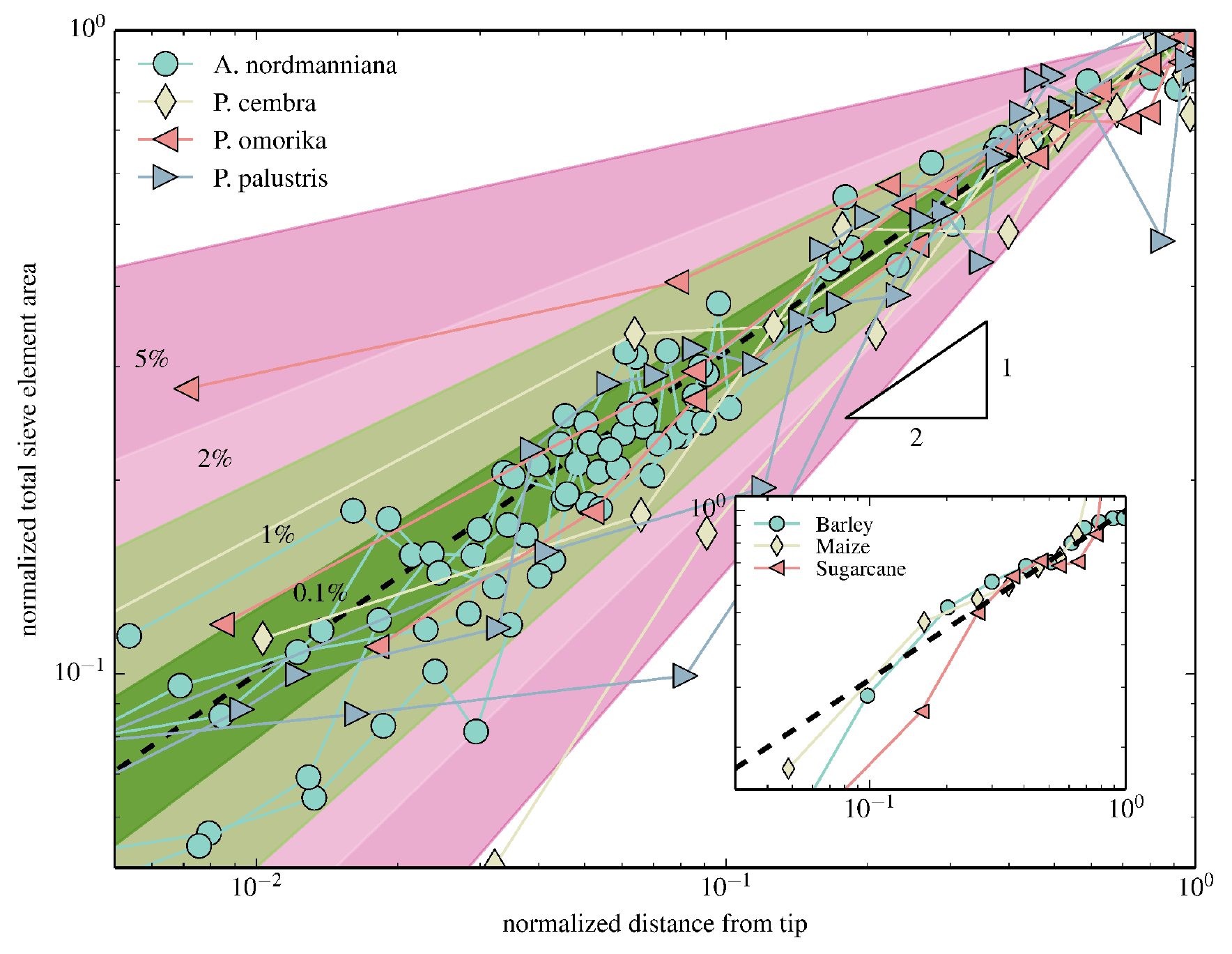}
\caption{(Colour online) Conifer needle phloem follows the $\alpha=1/2$ law. Phloem area $A/A(L)$ plotted as a function of distance from the needle tip $x/L$ on double-logarithmic axes. The dashed line has slope $1/2$. 
The colored regions correspond to all power laws 
$A_\alpha(x) \sim (x/L)^\alpha$ whose power dissipation exceeds that
of the optimal solution $A_{1/2}$ by the given percentages.
Most of the needles analyzed fall within the $1\%$ range $\alpha \in
[0.35, 0.65]$, see Table~1.
The inset shows phloem area data from the monocots barley, maize, and sugarcane obtained from \cite{Colbert1982,Russell1985,Dannenhoffer1990}. Barley is in good accord with the $\alpha=1/2$ scaling, while maize and sugarcane also only show approximate sub-linear dependences on distance from leaf tip.
 \label{fig:master-a-loglog}}
\end{figure}

For a given conductive phloem volume $V_0$, we can now determine the area distribution $A(x)$ which minimizes the viscous power dissipation. Using the method of Lagrange multipliers and the calculus of
variations, one finds that the distribution which
minimizes \eqref{eqn:functional} under the constraint $\int_0^L A(x)\,\mathrm d x=V_0$
is

\begin{align}
A(x) = \frac{3}{2} \frac{V_0}{L} \left( \frac{x}{L} \right)^{1/2}
 = \frac{3}{2} \langle A \rangle \left( \frac{x}{L} \right)^{1/2},\label{eq:A}
\end{align}
where $\langle A \rangle$ is the average total area of sieve elements. Assuming a bundle of sieve elements with constant cross sectional area,
this result may be translated immediately to total number of sieve elements by

\begin{align}
N(x) = \frac{A(x)}{A_0} = 
  \frac{3}{2} \langle N \rangle \left(\frac{x}{L}\right)^{1/2},\label{eq:N}
\end{align}
where $A_0$ is the cross sectional area of a single sieve element. From Eqns. ~\eqref{eq:A} and \eqref{eq:N}, we conclude that a scaling of phloem sieve element number or area with the power $\alpha = 1/2$ minimizes the viscous power dissipation. The observed scaling exponents (see Fig.~\ref{fig:master-a-loglog} and table 1) are close to these values, suggesting that the sieve element areas roughly follow the theoretical optimum. {We note that the square-root scaling in Eq.~}\eqref{eq:N}{ is identical to that found by Zwieniecki \textit{et al.}} \cite{Zwieniecki2006a} { for the tapering of pine needle xylem conduits. The coupling between water flow and phloem loading required to maintain a constant photosynthetic rate along the needle is thus responsible for 
driving this remarkable convergence in vascular architecture.} In the context of leaf development, we note that pine needles grow from a meristem located at the base of the needle which gradually propels the tip away from the growth zone. Newly formed tissue at the base of the needle gradually becomes mature and loses its ability to change its structure as the needle extends. The distribution of phloem conduits along the needle length thus appears to be either predetermined or rely on exchange of information between the tip of the needle and the meristem.

While our model is not generally applicable to complex reticulate or anastomosing vein networks, we expect it to be suitable for analysis of leaves with parallel veins, given that the assumption of constant sieve element properties is valid. Evidence to support this hypothesis is found in studies by R. F. Evert and co--workers \cite{Colbert1982, Russell1985, Dannenhoffer1990}, who observed similar trends in grasses. For example, phloem area in barley, maize, and sugarcane roughly follow the $\alpha = 1/2$ law, (Fig.~\ref{fig:master-a-loglog} inset), suggesting that the energy dissipation criterion leading to the prediction $\alpha=1/2$ is broadly applicable. We show in the appendix that the constant volume constraint imposed when obtaining $\alpha=1/2$ can be relaxed, and that sub-linear scalings is a general feature of the energy minimization principle. 

Previous works \cite{Jensen2011,Jensen2012}, which focused on one-dimensional models of flow in sieve elements, identified the transport velocity (phloem sap flux density) as an important physiological parameter. In fact, optimizing the sieve element radius ($r_0$ in eqn. \ref{eqn:net-currents}) for maximum flux at the whole plant level results in predictions that are in agreement with experimental observations \cite{Jensen2011}. Interestingly, we find (see Supplementary Information) that while the observed conduit distribution (i.e. $A\propto x^{1/2}$) minimizes the energetic cost of transport for a fixed tube radius $r_0$, it does not maximize the average flux density. The size of individual phloem cells at the level of the whole tree thus appears to be optimized for flux density, while the arrangement of tubes in the needle minimizes the energetic cost of transport, working in concert to produce
an efficient system of nutrient translocation.

\section{Discussion}
In this work, we studied the physical properties of nutrient transport
in the phloem of conifer needles. We measured the geometrical properties
of needle phloem in several conifer species, varying over one order of magnitude in length, and found that their cross sectional area distribution roughly follows the law $A(x) \sim x^{1/2}$.
We presented  a simple mathematical model which is able
to rationalize the observed needle tube geometry by 
means of minimization of the energy dissipated
during flow. Expenditure of energy is unavoidable since although
the transport is entirely passive by virtue of the osmotic flow process, the plant is 
forced to maintain an osmotic gradient, consuming energy in the process.
We found that experimental data from several species of conifers agree
well with the theoretically derived law of area distribution.

Simple models such as the ones considered in this work may not only
elucidate the properties of structures with modest complexity
we see in the living world, but also serve as an important stepping
stone to further understanding of more complicated systems. 
The basic underlying constraints and functional requirements that dictate needle design in conifers are not unique to this group of plants. Data from parallel-veined grasses indicate similar trends, and the design requirements are expected to hold for plants with reticulate venation patterns. The same mathematical model can potentially be extended to predict vascular distribution when the leaf lamina is broad and the single vein is replaced by an extensive reticulate network.

The conifer needle belongs to a general class of network systems that follow a principle of energy dissipation minimization. Other important members of this class which is not constrained to one-dimensional or even
planar systems include the networks of blood vessels in animals
\cite{Murray01031926xx,Murray01051926xx}, the xylem
vascular system in plants \cite{Katifori2010a,McCulloh2003xx} 
and even river basin networks in geomorphology \cite{rinaldo1992minimumxx},
thus establishing the importance of optimization considerations.

Finally, we point out that in recent years the constructal law,
stipulating that all living organisms are built so as to optimally
facilitate flow (of fluids, stresses, energy) has enjoyed some
success \cite{Bejan2011,Bejan2008} in explaining the structure and
apparent design of biological systems. The findings we report in this 
paper appear to be in accordance with the basic ideas 
from constructal theory.

\section{Materials and Methods}
Needles of mature \emph{Abies nordmanniana, Pinus palustris, Pinus cembra}, and \emph{Picea omorika }were collected in May and June of 2013. Samples of A. nordmanniana, P. cembra and P. omorika were taken in Denmark, while \emph{P. palustris} needles were collected in Florida (USA) and shipped to Denmark by courier.

Needles were embedded in low-melting point agarose (Sigma-Aldrich) and sectioned with a vibrating blade microtome (Leica Microsystems) to ensure uniform section thickness of 100 $\mathrm{\mu m}$. Sections were imaged using a confocal laser scanning microscope (SP5X, Leica Microsystems). In this way, 3 to 6 needles of average length (see Figs.~\ref{fig:abies-cross-section} and \ref{fig:needles}) from each species were analyzed. The number of sieve elements and their cross-sectional area were quantified using the image analysis software Volocity (Version 5.3, PerkinElmer).

By way of fluorescence staining with the live-cell marker carboxy fluorescein diacetate (Sigma-Aldrich) non-functional sieve elements were excluded. Needle sections of $5 \mathrm{mm}$ length were incubated in carboxy fluorescein for 15 minutes after which sections were made and analyzed under the microscope. For all species, only a few strongly deformed sieve elements at the abaxial side of the phloem bundle in which almost no cytoplasm was visible were found to be dead, i.e. non-functional. These cells were not included in the analysis.

\section*{Appendix}
\label{app:a} 
We consider a extension of the constant volume constraint 
by introducing a more general dependency on some power of the
total area,

\begin{align}
\int_0^L \mathrm d x\, A(x)^\gamma = K,
\end{align}
where we now think of $K$ as a general cost of building material
and metabolism which scales with cross-sectional area in a nonlinear way. Constraints of this type
have been used extensively in the field of complex distribution networks
\cite{Katifori2010a,bernot2009optimalxx}.
The optimization of Eq.~\eqref{eqn:functional} under this generalized
constraint predicts a scaling of
\begin{equation}
A(x) \sim x^{1/(\gamma + 1)}.
\end{equation}
We note that this result is robust: the optimal area scaling is sub-linear, whatever the value of the scaling power $\gamma$ of the cost function.    

\section*{Acknowledgements}
The work of Kaare H. Jensen is
supported by the Air Force Office of Scientific Research 
(Award No.: FA9550-09-1-0188), the National Science Fundation 
(Grant No.: DMR-0820484), the Danish Council for Independent Research $|$ Natural Sciences, and the Carlsberg Foundation.
The work of Henrik Ronellenfitsch is supported by the IMPRS for Physics of Biological and Complex Systems, G\"ottingen. Eleni Katifori acknowledges the support of the Burroughs Wellcome Fund through the BWF Career Award at the
Scientific Interface.

\section*{Data Accessibility}
The phloem geometry data will available digitally at \cite{dryad} at the 
time of publication.

\begin{table}
\centering
\begin{tabular*}{\hsize}
{@{\extracolsep{\fill}}llll}
species & scaling exponent $\alpha$ & $R^2$ & $p$ value \cr
\hline
Pinus palustris & & & \cr
\hline
& $0.51$ & $0.94$ & $<0.001^{***}$ \cr
& $0.60$ & $0.90$ & $<0.01^{**}$ \cr
& $0.67$ & $0.82$ & $<0.001^{***}$ \cr
\hline
average & $0.59 \pm 0.06$ & & \cr
\hline
Abies nordmanniana & & & \cr
\hline
& $0.38$ & $0.71$ & $<0.001^{***}$ \cr
& $0.65$ & $0.95$ & $<0.001^{***}$ \cr
& $0.72$ & $0.92$ & $<0.001^{***}$ \cr
& $0.48$ & $0.84$ & $<0.01^{**}$ \cr
& $0.48$ & $0.92$ & $<0.001^{***}$ \cr
& $0.65$ & $0.99$ & $<0.001^{***}$ \cr
\hline
average & $0.56 \pm 0.12$ & & \cr
\hline
Pinus cembra & & & \cr
\hline
& $0.73$ & $0.92$ & $<0.01^{**}$ \cr
& $0.41$ & $0.95$ & $<0.01^{**}$ \cr
& $0.61$ & $0.91$ & $<0.05^{*}$ \cr
\hline
average & $0.58 \pm 0.13$ & & \cr
\hline
Picea omorika & & & \cr
\hline
& $0.30$ & $0.92$ & $<0.001^{***}$ \cr
& $0.47$ & $0.98$ & $<0.001^{***}$ \cr
& $0.58$ & $0.99$ & $<0.001^{***}$ \cr
\hline
average & $0.45 \pm 0.11$ & & \cr
\hline
\end{tabular*}
\caption{Results of least-squares fitting total sieve tube area 
$\log A(x) = c + \alpha \log x$ for individual needles.
Asterisks highlight significance levels (${}^{***}:p<0.001$, 
${}^{**}:p<0.01$, ${}^{*}:p<0.05$), error estimates given correspond
to one standard deviation.
\label{tab:fits}}
\end{table}


\end{document}